\documentclass[aps,prb,reprint,preprintnumbers,superscriptaddress,amsmath,amssymb,bibnotes,longbibliography]{revtex4-2}

\usepackage{graphicx}
\usepackage{dcolumn}
\usepackage{bm}
\usepackage[colorlinks,linkcolor=blue,anchorcolor=blue,citecolor=blue,urlcolor=blue,filecolor=blue,menucolor=blue,runcolor=blue]{hyperref}
\usepackage{float}
\usepackage{ulem}

\begin{document}

	\title{ Multigap nodeless superconductivity in Dirac semimetal PdTe}
	
	\author{Fengrui Shi}
	\author{Weilong Qiu}
	 \affiliation  {Center for Correlated Matter, School of Physics, Zhejiang University, Hangzhou 310058, China}
    \author{Chufan Chen}
	\email[Corresponding author: ]{cf.chen@nus.edu.sg}
	\affiliation  {Center for Correlated Matter, School of Physics, Zhejiang University, Hangzhou 310058, China}
	\affiliation  {Department of Electrical and Computer Engineering, National University of Singapore, Singapore 117583, Singapore}
	\author{Chunqiang Xu}
  \affiliation  {School of Physical Science and Technology, Ningbo University, Ningbo 315211, China}
    \author{Yan Zhang}
    \author{Hao Zheng}
    \author{Yuwei Zhou}
	\author{Dongting Zhang}
	\author{Mengwei Xie}
	\affiliation {Center for Correlated Matter, School of Physics, Zhejiang University, Hangzhou 310058, China}
	\author{Huiqiu Yuan}
    \affiliation {Center for Correlated Matter, School of Physics, Zhejiang University, Hangzhou 310058, China}
		\affiliation {State Key Laboratory of Silicon Materials, Zhejiang University, Hangzhou 310058, China}
	\author{Shiyan Li}
	\affiliation {State Key Laboratory of Surface Physics, and Department of Physics, Fudan University, Shanghai 200438, China}

  \author{Yang Liu}
	\author{Chao Cao}
	\affiliation {Center for Correlated Matter, School of Physics, Zhejiang University, Hangzhou 310058, China}
	
	\author{Xiaofeng Xu}
	\affiliation{School of Physics, Zhejiang University of Technology, Hangzhou 310023, China}

	\author{Xin Lu}
	\email[Corresponding author: ]{xinluphy@zju.edu.cn}
	\affiliation  {Center for Correlated Matter, School of Physics, Zhejiang University, Hangzhou 310058, China}
	
	\begin{abstract}
		PdTe has recently been reported to be a type-II Dirac semimetal while a bulk nodal and surface nodeless superconductivity (SC) has been claimed to coexist. In this work, we applied point-contact spectroscopy (PCS) method to systematically study the superconducting gap in PdTe single crystals with a SC transition temperature $T_{c}=4.3$ K. The obtained differential conductance curves show a common deviation from a single-gap superconducting behavior and can be better fitted by a two-gap Blonder-Tinkham-Klapwijk model, suggesting the larger gap $\Delta_{L}$ with $2\Delta_{L}$=3.7 $k_{B}T_{c}$ and the smaller gap $\Delta_S$ yielding $2\Delta_{S}$=1.1-2.2 $k_{B}T_{c}$ with a weak interband scattering. The variations of conductance spectra among different contacts are proposed to be caused by the anisotropy of Fermi surface topology associated with different gaps.
	\end{abstract}
	
	\maketitle
	
	\section{Introduction}
	
	Topological superconductivity (TSC) has attracted significant attention due to its fascinating physical properties \cite{TP1,TP2,TP3,TP4,TP5} and potential applications in fault-tolerant quantum computation \cite{quamtumcomputation1, quamtumcomputation2, quamtumcomputation3}. In TSCs, Majorana fermions are expected to appear on the surface or edge of TSCs,  serving as an active area of research \cite{p1,p2,p3,p4,p5,p6}. There are typically two main approaches to achieve TSC: One promising approach is to construct topological semimetal (insulator)-superconductor heterojunctions, by inducing SC in topological materials via proximity effect, as demonstrated in Bi$_{2}$Se$_{3}$/ NbSe$_{2}$ \cite{Bi2Se3}, Bi$_{2}$Te$_{3}$/ NbSe$_{2}$ \cite{Bi2Te3} and TlBiSe$_{2}$ film/ Pb $(111)$ \cite{TI}. An alternative strategy to search for TSC is to explore intrinsic superconductors with nontrivial topological bands, such as PdTe$_{2}$ \cite{PdTe2}, PbTaSe$_{2}$ \cite{PbTaSe2}, Bi$_{2}$Se$_{3}$ \cite{Bi2Se3PCS}, Cu$_{x}$Bi$_{2}$Se$_{3}$ \cite{CuxBi2Se31}, and Cd$_{3}$As$_{2}$ \cite{Cd3As21,Cd3As22}.

	Among them, PdTe, a type-\uppercase\expandafter{\romannumeral2} Dirac semimetal \cite{sampleJinrongying,sampleindia,CHEN201623} with a hexagonal NiAs-type structure and superconducting transition temperature around 4.3--4.6 K, has attracted great interest due to the possible coexistence of nodal bulk and nodeless surface superconductivity observed in the angle resolved photoemission spectroscopy (ARPES) measurement \cite{Hassan}. Its nontrivial topological nature has been further confirmed by band-structure calculations and de Haas–van Alphen (dHvA) measurements \cite{CHEN201623,anisotropy,2023Jinrongying}. However, experimental results on the nature of superconducting pairing gap are still under debate:  nodal superconducting order parameter has been argued based on heat capacity measurements, where the electronic heat capacity $C_{e}$ shows a linear relation to $T^3$ at the temperature range of $T_c/3$ $<$  $T$ $<$ $T_{c}$ \cite{2023Jinrongying}. Additionally, an ultralow-temperature heat capacity fits better with \textit{p}-wave symmetry  rather than \textit{s}- or \textit{d}-wave symmetry \cite{pwaveheatcapacity}. In contrast, ultralow-temperature thermal conductivity results support a nodeless and multiband superconducting gap structure \cite{Lishiyan}, consistent with another heat capacity and upper critical field measurements \cite{anisotropy}. Therefore, a more detailed study with different techniques is desirable to probe the superconducting properties in PdTe and point-contact spectroscopy serves as a powerful tool to determine its SC gap structure.

	In this study, both soft and mechanical point-contact spectroscopy (SPCS and MPCS) were applied to investigate the superconducting gap in PdTe single crystals. Most of the differential conductance curves deviate from the single-gap s-wave fitting and should be better described by a two-gap Blonder-Tinkham-Klapwijk (BTK) model. Detailed analysis supports the larger gap $\Delta_{L}$=0.55-0.6 meV and the smaller one $\Delta_{S}$=0.2-0.4 meV, while their temperature dependence suggests a weak interband scattering. The observed variation of conductance spectra and a broad distribution of the smaller gap $\Delta_{S}$ are attributed to the Fermi surface anisotropy for respective gap bands.

	\section{Method}
	
	High-quality PdTe single crystals were grown by the method described in Ref. \cite{2023Jinrongying} and all screened samples have similar superconducting behaviors, ensuring  reliable results. Figures 1 (a) and (b) show the temperature-dependent electrical resistance and special heat of freshly cleaved PdTe single crystals from the same batch. Both measurements exhibit a sharp superconducting transition at $T_{c}=$4.3 K, confirming the high quality of crystals. 
	
	\begin{figure}
		\includegraphics[angle=0,width=0.49\textwidth]{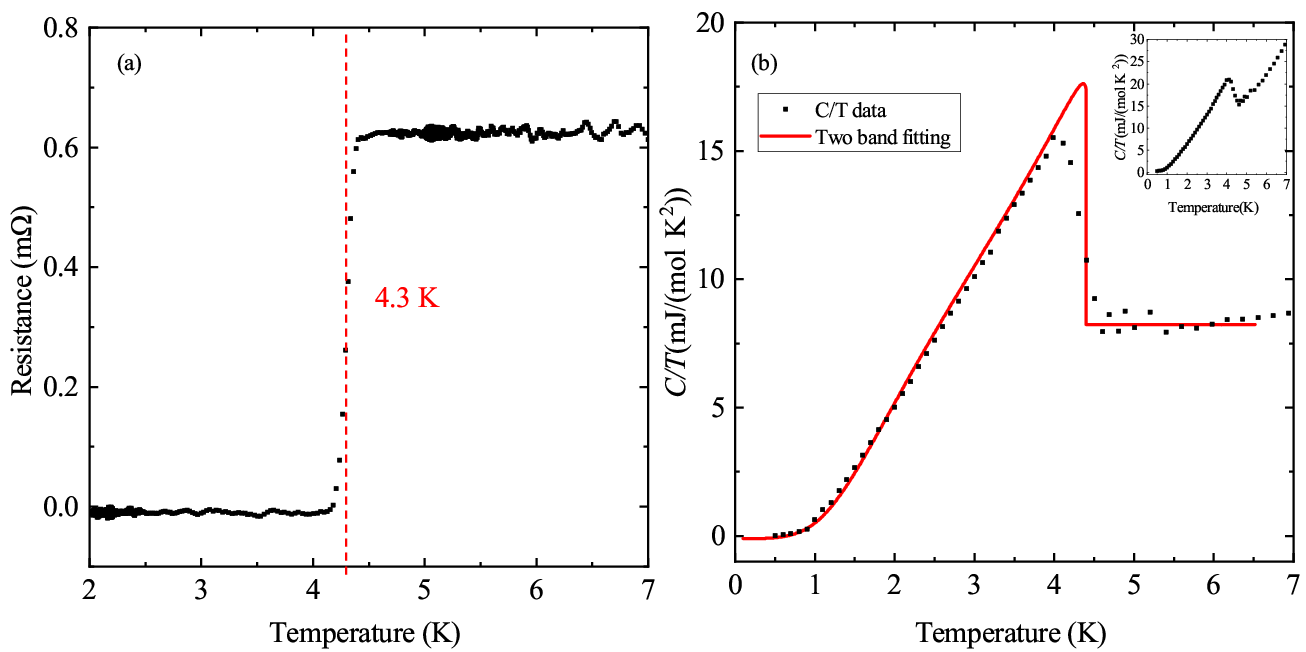}
	\vspace{-12pt} \caption{\label{Figure1}(color online) (a) Resistance measurement of single-crystal PdTe. (b) Electronic specific heat capacity of PdTe and the fitting curves are shown in black dots and red line, respectively. The insert shows the original specific heat capacity data. The dashed line in (a) indicates a sharp superconducting transition $T_{c}$ $=$ 4.3 K.}
		\vspace{-12pt}
	\end{figure}
	
Point-contact spectroscopy is a powerful tool to probe the pairing symmetry of superconductors, based on the Andreev reflection process at the normal metal and superconductor interface. In this work, we employed both soft and mechanical point-contact techniques to characterize our PdTe samples: The soft point-contact spectroscopy was performed by attaching a 30-$\mu$m-diameter Au wire to the sample surface with a drop of silver paint at the end, where thousands of parallel contact channels were formed by individual silver particles. Mechanical point contacts were employed in a needle-sample configuration, where the distance between the sharp Au tip and sample can be accurately manipulated by pizeoelectric nanopositioners. Compared to silver paint, the tip-sample contact in mechanical point contacts offers greater flexibility and a significantly smaller contact area. In both cases, a “single” local contact was used to probe the local electronic structure beneath the effective electrical contact area in both MPCS and SPCS configurations.
	
	The PCS differential conductance as a function of bias voltage $G(V)$ was recorded by the conventional lock-in technique in a quasi-four-probe configuration. In order to ensure a clean contact, the sample surfaces for point contacts are either from a clean crystal or obtained on freshly cleaved or broken crystals. An Oxford Instruments cryostat with a ${^3}$He insert (base temperature 0.3 K) was used for our standard four-probe electrical resistance and PCS measurements and our specific heat data were measured in a Quantum Design PPMS-8T with a ${^3}$He insert. We note that silver paint or gold tip used for point contact was directly applied to the bare or fresh surface from the broken PdTe crystal. Laue diffraction or x-ray diffraction (XRD) was utilized to analyze the crystal orientation, and samples were confirmed to have random directions, which is difficult to prepare in the major symmetric axis due to limited sample size.

	\section{Results}
	
	Dozens of soft and mechanical point contacts have been achieved on PdTe single crystals to investigate its superconducting gap and Fig. 2 shows a representative set of differential conductance curves for contacts on various sample surfaces (bare or freshly cleaved) at the lowest temperature $T$= 0.3 K. They all exhibit typical double peaks in the conductance due to Andreev reflection but with additional shoulder or wiggle features.  A single-gap s-wave BTK fitting (black lines) obviously fails to reproduce the experimental data, as evidenced by the pronounced shoulders and extra conductance near zero-bias, supporting the existence of multi-gaps. Therefore, we have considered a two-gap s-wave BTK model with two different gap sizes in PdTe, and the conductance can be simulated as G(V)=$\omega_L$G$_1$(V)+(1-$\omega_L$)G$_2$(V) (0$\leq$ $\omega_L$ $\leq$ 1), where the two components correspond to the contributions from two separate gaps and the parameter $\omega_{L}$ denotes the spectra weight from the larger gap $\Delta_{L}$. For an optimal fitting by the two-gap BTK model, a larger gap around 0.55-0.6 meV and a smaller one around 0.2-0.4 meV can be consistently obtained as shown in Fig. 2.
	
	\begin{figure}
	\includegraphics[angle=0,width=0.49\textwidth]{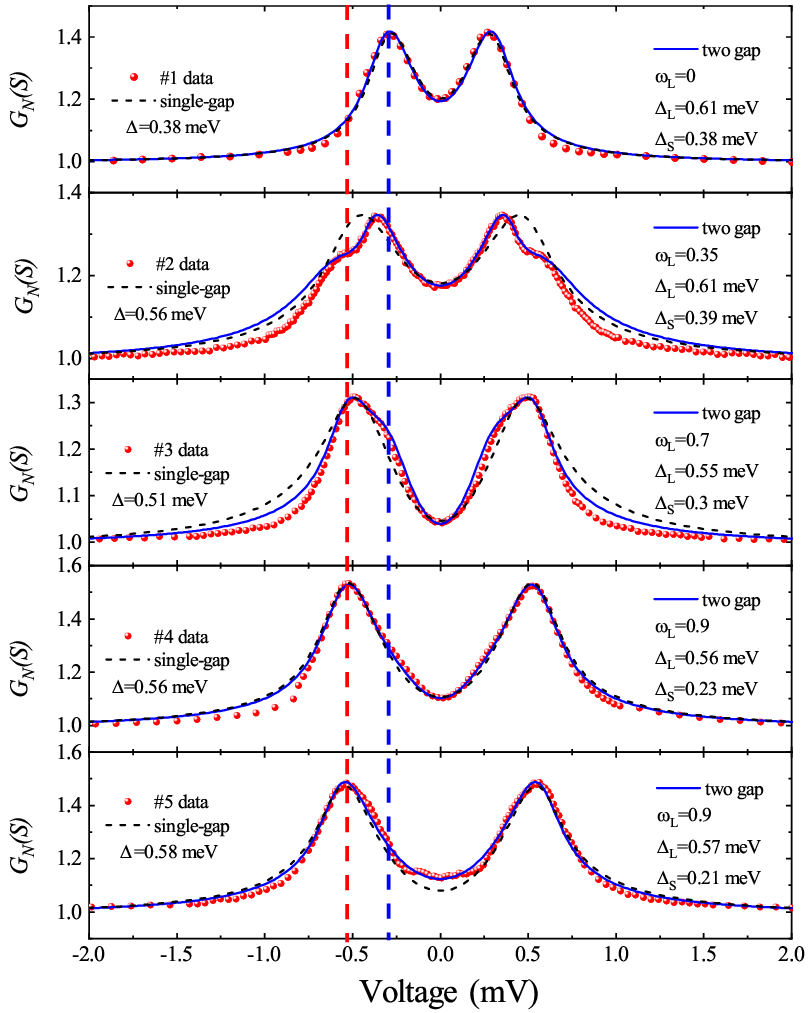}
		\vspace{-12pt} \caption{\label{Figure2}(color online) Normalized PCS differential conductance curves measured at different point contacts. The single-gap and two-gap BTK fitting curves are shown by the black and blue lines, respectively. The red and blue dashed lines are guide to the eye to the existence of two gaps.}
		\vspace{-12pt}
  \end{figure}
	
	To confirm the multigap nature of superconducting PdTe, we have also analyzed the specific heat data in Fig. 1(b), which shows the temperature dependence of electronic specific heat at low temperatures, $C_{e}$/$T$ versus $T$, whereas the total heat capacity data are presented in the inset of Fig. 1(b). The normal-state heat capacity of PdTe can be well described by the formula $C/T=\gamma_{n} +\alpha T^{2} +\beta T^{4}$, where the electronic heat capacity coefficient $\gamma_{n}$ =8.49 mJ/(mol*K$^{2})$ and the phonon contribution coefficient $\alpha$ =0.27 mJ/(mol*K$^{4})$, $\beta$= 0.00304 mJ/(mol*K$^{6})$. The electronic specific heat $C_{e}$ in the superconducting state can be extracted by the formula $C_{e}/T=C/T- \alpha T^{2} -\beta T^{4}$. Similarly, a two-gap s-wave model has to be applied to fit the heat capacity data with $C_{e}=\omega_L C_{e\Delta_{L}}+ (1-\omega_L) C_{e\Delta_{S}}$, where $\omega_L$ represents the $\Delta_{L}$ contribution to the heat capacity. As shown  in Fig. 1(b), the best fitting to the specific heat data yields  $2\Delta_{L}$=3.7 $k_{B}T_{c}$ and $2\Delta_{S}$=1.7 $k_{B}T_{c}$, where $\Delta_{L}$ contributes the majority of the specific heat $C_e$ ($\omega_L \sim$ 90-95\%). These results are consistent with our point-contact measurements, confirming the multiband nature of the superconductivity in PdTe.

	In order to characterize the temperature evolution of the superconducting gaps, we have selected contact \#2 in Fig. 2 due to its notable shoulder feature in G(V) and plot its temperature dependence in Fig. 3(a). As the temperature increases, the double peaks gradually get smeared into a broad zero-bias peak, and eventually disappear at the same SC transition temperature $T_{c}$=4.3 K as the specific heat $T_{c}$. As illustrated in Fig. 3(a), the conductance curves G(V) can be well fitted by the two-gap s-wave BTK model, while the broad zero-bias peak at high temperatures makes it unreliable to extract $\Delta_{L}$ and $\Delta_{S}$, respectively. The obtained larger gap $\Delta_{L}$ follows a standard BCS temperature behavior as in Fig. 3(c), yielding $\Delta_{L}$=0.61 meV and 2$\Delta_{L}$/k$_{B}$T$_{c}$=3.54 roughly in the weak-coupling limit. In contrast, the smaller gap $\Delta_{S}$=0.4 meV deviates from the BCS behavior, and seems to follow a BCS behavior with an assumed $T_c \sim$ 2.7 K initially but persists up to the authentic $T_c \sim$ 4.3 K with a long tail, possibly due to a weak interband scattering as discussed later \cite{1theory}. 
	
		\begin{figure}
		\includegraphics[angle=0,width=0.49\textwidth]{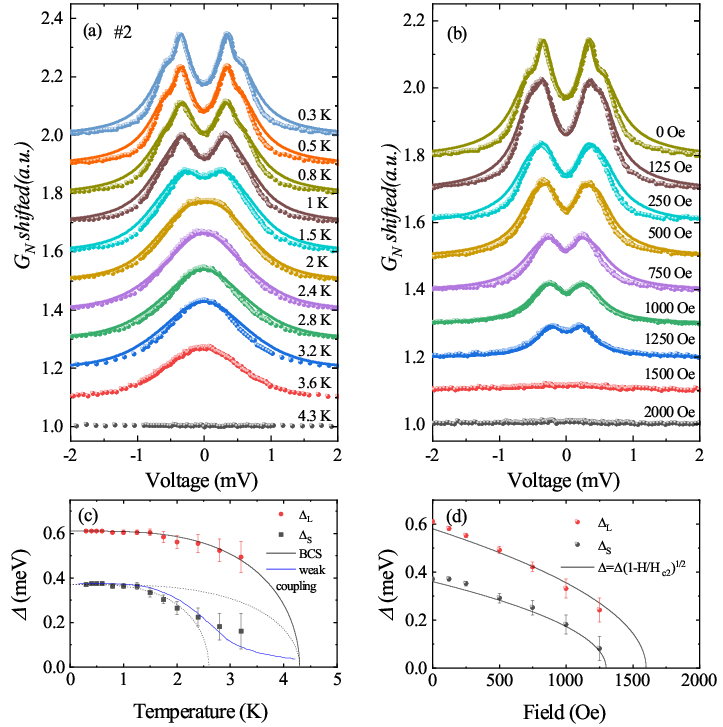}
		\vspace{-12pt} \caption{\label{Figure3}(color online) Differential conductance curves for point contact \#2 on PdTe as a function of (a) temperature and (b) magnetic field at 0.3 K. The solid lines represent the two-gap s-wave BTK fitting curves. Two gaps extracted from the fittings in (a) and (b) are shown in (c) and (d), respectively. The black lines represent the BCS behavior with respect to temperature and field, while the blue line corresponds to the weak interband scattering fitting, as described in Ref. \cite{1theory}}.
		\vspace{-12pt} 
		\end{figure}

Figure 3(b) shows the magnetic-field dependence of G(V) curves for contact \#2 on PdTe at 0.3 K along with the two-gap s-wave fitting. As the field is ramped up to 2000 Oe, the superconducting signals are gradually suppressed, where the height of double peaks are reduced and the peak positions shift toward zero-bias voltage. The extracted gap values as a function of field are shown in Fig. 3(d), complying with the typical type-II superconductor behavior with $\Delta (H)=\Delta _{0} \sqrt{1-H/H_{C2} }$.

		\begin{figure}
		\includegraphics[angle=0,width=0.49\textwidth]{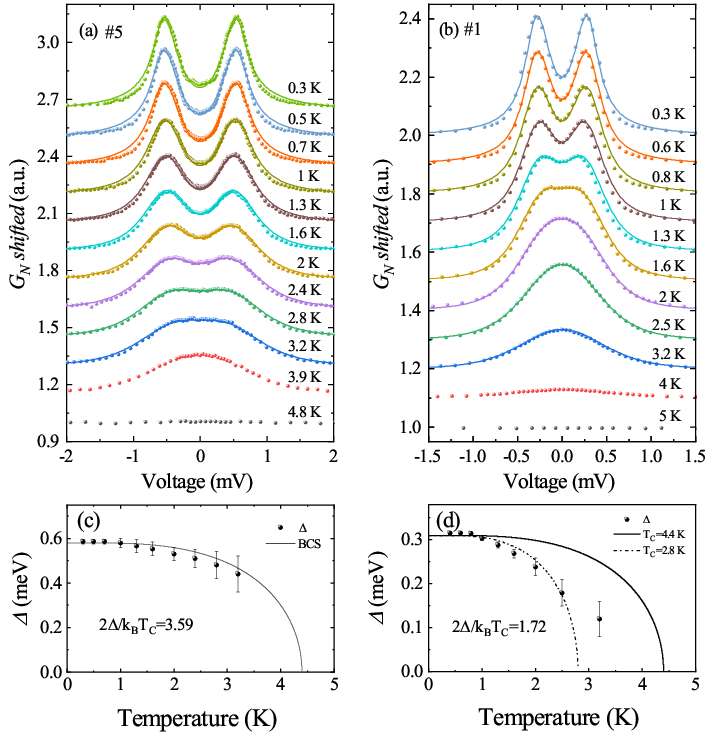}
		\vspace{-12pt} \caption{\label{Figure4}(color online) The normalized differential conductance curves for contacts \#5 and \#1 are shown in (a) and (b), only with a dominant larger or smaller gap feature, respectively. The solid lines represent fits of the single-gap s-wave BTK model to the data. The respective gaps obtained from the fittings in (a) and (b) are shown in (c) and (d), in comparison with the BCS curves in solid, while the dashed line in (d) shows a BCS curve with a T$_c \sim$ 2.8 K.
		}
		\vspace{-12pt}
	\end{figure}

	While contact \#2 in Fig. 3 has displayed notable spectra weight from both gaps $\Delta_{L}$ and $\Delta_{S}$ in the conductance curves G(V) , we focus on another two sets of PCS data with typical double peaks for more detailed analysis, where the spectra weight $\omega_L$ for the larger gap is in the opposite limit of 90\% for contact \#5 or zero for contact \#1, respectively. In such a case, we can just fit the data with a single-gap BTK model and double-check the temperature dependence of both superconducting gaps. Fig. 4 (a) and (b) show the temperature-dependent G(V) curves for two contacts, respectively, where double peaks gradually merge into a single broad zero-bias peak and disappear for temperatures above T$_{c}$ and the absence of dips in G(V) ensures the ballistic nature of both contacts. For Fig. 4(a), a single-gap BTK model can indeed have a perfect fitting with $\Delta=\Delta_{L}$ and the obtained $\Delta_{L}$ values follow the same BCS temperature behavior as that observed in Fig. 2(c). Similarly, the single gap fittings with $\Delta=\Delta_{S}$ are illustrated in Fig. 4(b), and $\Delta_{S}$ fails to follow the standard BCS temperature dependence also and retains a long tail till $T_c$, exactly the same as in Fig. 2(c). Considering the consistency between different contacts, we would argue that the larger gap $\Delta_{L}$ follows the BCS temperature behavior while the smaller one $\Delta_{S}$ deviates from the BCS behavior with a long tail, even though both gaps close at the same SC transition temperature $T_{c}$ for PdTe. Such a behavior may suggest a weak interband scattering between two gaps, where $\Delta_{S}$ value can be raised by $\Delta_{L}$ and its SC transition temperature can be elevated to the same value as $\Delta_{L}$ but with a long tail. These behaviors are consistent with a weak interband scattering theory \cite{1theory}, where $\Delta_{S}$ will be slightly enhanced, while its temperature evolution deviates from the standard BCS curve, exhibiting a long tail terminating at the $T_c$ of $\Delta_{L}$ as in Fig. 3(c) and 4(d). In the extreme case without any interband coupling, the two gaps will open at different transition temperatures, while the presence of interband scattering will yield a common critical temperature instead. 
	
		\begin{figure}
		\includegraphics[angle=0,width=0.49\textwidth]{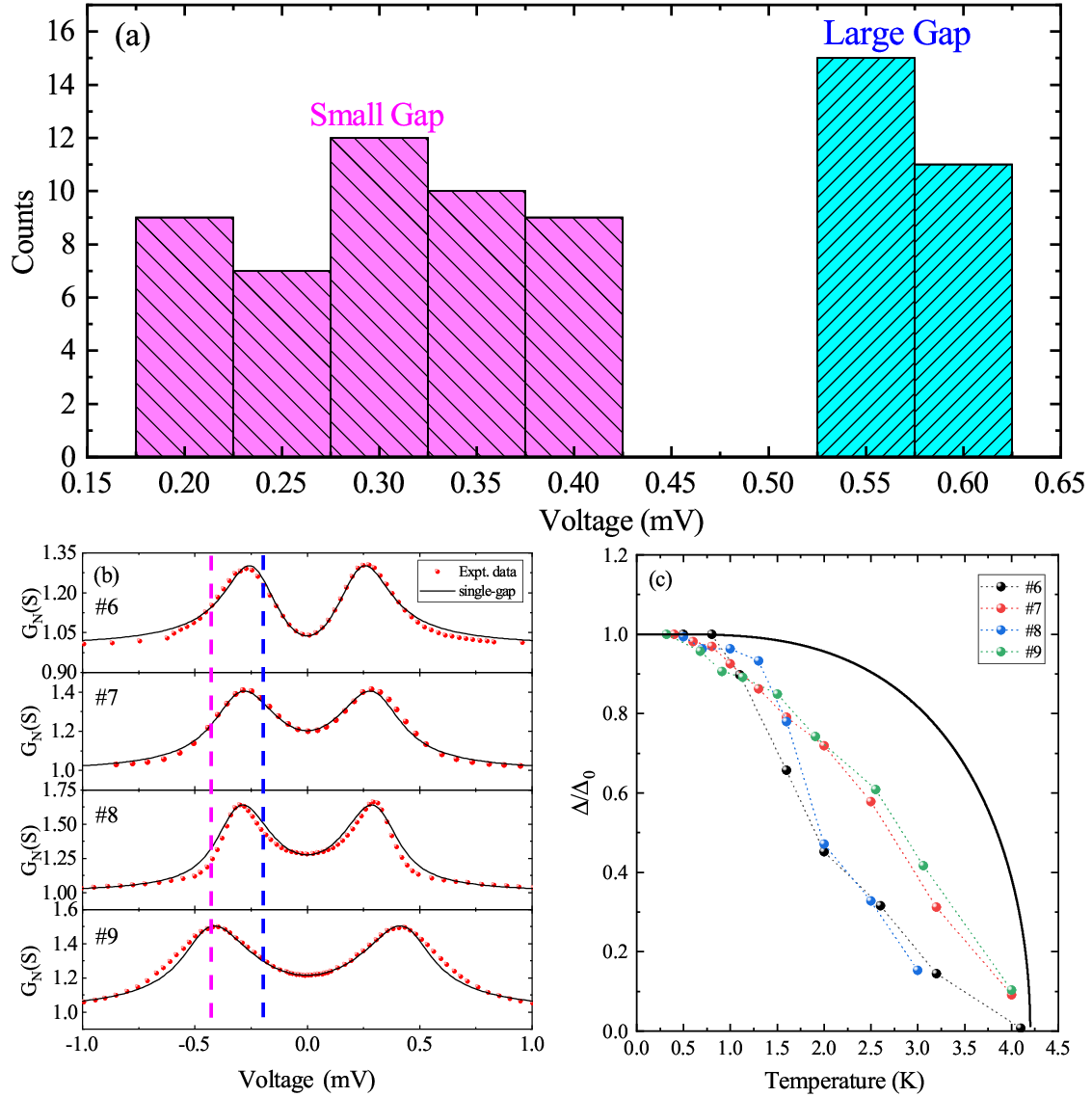}
		\vspace{-12pt} \caption{\label{Figure5}(color online) (a) The histogram plot of $\Delta_{L}$ and $\Delta_{S}$ observed for PdTe in dozens of point contacts. (b) Four representative G(V) curves indicate a broad distribution of $\Delta_{S}$ in comparison with its single-gap BTK fitting. Magenta and blue dashed lines are guides to the eye, defining the distribution of $\Delta_{S}$.	(c) Normalized gap values for the selected point contacts \#6-\#9 show a consistent temperature dependence, deviating from the BCS behavior (black line).}
		\vspace{-12pt}
	\end{figure}
	
	In order to further confirm the existence of both $\Delta_{L}$ and $\Delta_{S}$, the histogram of obtained gap values is plotted in Fig. 5(a), illustrating the statistical distribution of $\Delta_{S}$ and $\Delta_{L}$, respectively. The larger gap $\Delta_{L}$ is concentrated in a narrow range around 0.6 meV, while the smaller gap $\Delta_{S}$ owns a broad distribution between 0.2 and 0.4 meV. Four representative G(V) curves from mechanical PCS are shown in Fig. 5(b) in the absence of the larger gap $\Delta_{L}$. The $\Delta_{S}$ gap value can be obtained by optimally fitting the G(V) curves with a single-gap BTK model, ranging from 0.2 to 0.4 meV. If we track its temperature dependence, the normalized gap $\Delta_{S}/\Delta_{S}^{0}$ for all contacts follow the same temperature evolution as shown in Fig. 5(c), similar to contact \#1 in Fig. 4(d), and deviates from the standard BCS behavior, whereas $\Delta_{L}$ aligns well with the BCS curve as demonstrated by contact \#5 in Fig. 4(c). We thus conclude that the smaller gaps $\Delta_{S}$ within the range of 0.2-0.4 meV should share a common origin.

	\section{Discussion}
	
Our point-contact measurements have shown a clear presence of multiband SC in PdTe as discussed above and at least two-gap BTK fitting is required to capture the conductance features. We note that multigap SC in PdTe has been reported in previous studies. The band structure of PdTe investigated by ARPES and dHvA measurements \cite{2023Jinrongying, anisotropy} is consistent with theoretical calculations \cite{CHEN201623}. These results show that at least three bands cross the Fermi surface, favoring multiband superconductivity.

From our PCS results, the smaller gap $\Delta_{S}$ can be consistently observed in the G(V) curves, whereas the feature from the larger gap $\Delta_{L}$ can sometimes be vague and its spectra weight $\omega$ varies from 0$\%$ to 90$\%$ among different contacts. Such a variation of point-contact spectra has also been observed in the well-known multiband superconductor MgB$_{2}$ \cite{MgB2, MgB21, MgB22}: When the point-contact current is applied along the c axis of MgB$_{2}$, a pair of perfect double peaks are observed, signaling the smaller gap from the three-dimensional (3D) $\pi$ band. However, when the contact is in the $\it{ab}$-plane, four peaks or double peaks with clear shoulders at higher bias are reported, implying the presence of a larger gap owing to the two-dimensional (2D) $\sigma$ bands besides the $\pi$ band. In such a case, a 2D Fermi surface band may not always contribute to the conductance spectra, particularly when the contact direction is normal to the 2D surface, yielding distinct spectra shapes with directional PCS. If the larger gap $\Delta_{L}$ in PdTe also originates from a 2D band while $\Delta_{S}$ is associated with a 3D band, such a scenario could naturally explain the occasional absence of $\Delta_{L}$ while $\Delta_{S}$  is persistently observed in the G(V) curves for PdTe. Since the crystals are randomly oriented in our PCS, we can not unambiguously identify the direction of the 2D band for $\Delta_L$ in PdTe from experiment so far.

To double-check our arguments, we have performed similar band-structure calculations and considered the density of states (DOS) of different bands in PdTe. Our obtained band structures are perfectly consistent with previous reports \cite{2023Jinrongying, CHEN201623}, where four bands, $\alpha$, $\beta$, $\gamma$ and $\eta$, cross the Fermi energy (details of the band-structure and DOS calculations can be found in the Supplemental Material). Our calculations support that the $\alpha$ band in PdTe shows a quasi-two-dimensional nature, reminiscent of the $\sigma$ band in MgB$_{2}$ \cite{MgB2band}. In addition, DOS calculations agree that the $\alpha$ band dominates the Fermi surface and contributes approximately 72\% of the total DOS, while the $\eta$ band accounts for 24\% and other two bands are negligible. These results are compatible with our speculations that the larger gap $\Delta_{L}$ may originate from the $\alpha$ band in PdTe, since the $\Delta_{L}$ has a major contribution to the heat capacity and exhibits a quasi-2D characteristic. On the other hand, the $\Delta_{S}$ can be attributed to the $\eta$ band with a 3D nature. Further careful investigations are desired to confirm that $\Delta_{L}$ is likely to originate from the 2D $\alpha$ band and $\Delta_{S}$ is from the 3D $\eta$ band. 

Moreover, for point-contact spectra where $\Delta_{L}$ is absent, scattered gap values for $\Delta_{S}$ have been observed as in Fig. 5, ranging from 0.2 to 0.4 meV with an even distribution. We have argued they should originate from the same band due to their normalized temperature dependence as in Fig. 5(c), and can be attributed to the 3D $\eta$ band. A broad distribution of $\Delta_{S}$ gap values from PCS may suggest an anisotropic SC gap with a minimum value of 0.2 meV for the 3D $\eta$ band in PdTe, where PCS in various directions would probe different gap values for $\Delta_{S}$. Additional spectra to prove the broad distribution of the smaller gap $\Delta_{S}$ due to Fermi surface anisotropy are included in the Supplemental Material and refer to a “smaller gap distribution” session for more detailed discussion. Notably, previous studies on PdTe have reported the existence of superconducting gap nodes from heat capacity and ARPES measurements. However, our point-contact results seem to favor a nodeless multigap superconductivity, and further experiments are necessary to explore detailed nature of $\Delta_{S}$. 


	\section{Summary}
In summary, we have employed both mechanical and soft point-contact spectroscopy to investigate the superconducting gap of the type-II Dirac semimetal PdTe crystals and our results strongly support a multiband superconductivity in PdTe with $2\Delta_{L}$=3.7 $k_{B}T_{c}$ and $2\Delta_{S}$=1.1-2.2 $k_{B}T_{c}$, accompanied by a weak interband scattering. We proposed that the observed variations of contact spectra can be ascribed to the Fermi surface anisotropy in PdTe, leading to the occasional absence of $\Delta_{L}$ and a scattered distribution of $\Delta_{S}$ in PCS G(V) curves. Further detailed studies, especially on better crystals with a clear identification of crystal orientations, are desired to confirm the correlation between the superconducting gap structure and Fermi surface topology.

	\section{ACKNOWLEDGMENT}
	Our work was supported by the National Key R\&D Program of China (Grant No. 2022YFA1402200), the National Natural Science Foundation of China (Grants No. 12174333, No. 12574147, No. 12274369, and No. 12304071) and the Key R\&D Program of Zhejiang Province, China (Grants No. 2021C01002). 
			
	\section{DATA AVAILABILITY}
	The data that support the findings of this article are openly available \cite{shi_2025_17837859}.


%

\end{document}